# Evolution of the problem « Search for life in the Universe » from some examples

Danielle Briot
*(Observatoire de Paris)*

Abstract
On the occasion of the 50th anniversary of the Drake formula, it appears interesting to briefly review the history of Astrobiology from the origins up to the epoch of the Drake formula. After recalling the main steps of this history during Antiquity, the Middle Ages, and the Renaissance, we point out some little known or unknown studies during the Modern and Contemporary epochs. Then we review the importance of astrobiology and the search for life in the Universe in the scientific publications during the 20th century.

Keywords :
History, Astrobiology

## 1. Some preliminary remarks

The science of Astrobiology presents at least two specific features. The first point is that this science can be named *Astrobiology, Exobiology, Bioastronomy,* and sometimes *Cosmobiology*, that is to say at least three or four words exist to indicate only one science. Actually, synonyms are very rare in the language of science where one word corresponds to only one object, and vice versa. There is a one-to-one relationship between the word and the object. The second point is that astrobiology is a science supposed to be the study of life in the Universe, but the first object of this study is to research whether this life does exist, and to discover it. Only if and when this first step is taken can the study of extraterrestrial life be done. Nowadays, this science is searching for itself. The object of the research is the search for the object.

For the fiftieth anniversary of the Drake formula, as we study the evolution of our knowledge since then, it is also interesting to study how the basis for approaches to the problem of the life in the Universe has evolved from the beginning to the Drake formula. However, the problem of the search for life in the Universe, connected to the problem of the Plurality of Worlds, has been studied so much (see for example Dick 1982 and Crowe 2008), and for so many centuries that we can make only a limited study, from a particular point of view, or using little known sources. In the present paper, we choose to talk about some quite unknown studies in addition to fundamental ones.

The problem of life in the Universe over the centuries, concerns the existence of other worlds capable of hosting life, or the search for life in the solar system, and also the definition of this life and thus the definition of our form of life, and its conditions.

It is also necessary to find out what some words which are familiar to us actually mean, but whose meaning and implications necessarily changed over the centuries.

Some existing writings are purely scientific, in the present meaning of this word, but others can be philosophical, religious, or can be legends or tales, more or less related to what is called science fiction today.

We first recall some significant steps in the search for other worlds, since Antiquity. For the modern and contemporary periods, we talk about some practically unknown studies. In the last part of this paper we study the use and the importance of the word astrobiology and its synonyms in astronomical literature.

## 2. Antiquity

The question of plurality of worlds dates back to classical Antiquity. However it is necessary to clarify the meanings of this expression. With modern scientific knowledge and the current mode of thinking, there is a tendency to think that *plurality of worlds* means the existence of planetary systems around other stars than our Sun. Now, let us not forget that the image of the world was very different from ours, and that stars were not at all considered as being other Suns. Plurality of worlds implied the existence, outside of our visible world which contains all known celestial objects, of other worlds similar to ours, maybe an infinite number of those, and unreachable.

The oldest known references concerning other worlds are those of the Greek philosophers Democritus (470 - 365 BC) and Epicurus (342 - 270 BC)

"*Worlds are unlimited and of different sizes. In some worlds there is no Sun and Moon, in others, they are larger than in our world, and in others more numerous. … Intervals between worlds are unequal. In some parts there are more worlds, in others fewer; some are increasing, some at their height, some decreasing; in some parts they are arising, in others failing …. There are some worlds devoid of living creatures or plants or any moisture.*"
Democritus

"*There is an infinity of worlds both like and unlike our world. For the atoms being infinite in number, as was already proved, are found  far out into space. For those atoms which are of such nature that a world could be created by them or made by them, have not been used up either on one world or a limited number of worlds… So that there nowhere exists an obstacle to an infinite number of worlds.*"  Epicurus, disciple of Democritus, "*Letter to Herodotus*"

Plato (428 - 348) and Aristotle (384 - 322 BC) took an opposite view:

"*…It follows that the world would be unique. There cannot be several worlds.*" Aristotle

## 3. Christian Era

From the Middle Ages another point needs to be taken into account: the *agreement?* between the theory of plurality of worlds and religion as taught. Two theological interpretations exist. On the one hand, the biblical story of the creation of the world in Genesis speaks of only one world. So occurrence of other worlds cannot happen. On the other hand, it is not specified that God did not create some other worlds and the almighty power of God implies that he could create other worlds. We have to note that this is not scientific reasoning, in the current meaning.

Saint Augustine (354 - 430 AD) was one of the first Christian authors to discuss the theory of plurality of worlds, which he had opposed, as did most medieval authors, including Albert the Great (1193-1280 AD) and Thomas Aquinas (1225 - 1274 AD).

The question of other worlds became a subject of controversy. In 1277, the bishop of Paris condemned 219 beliefs held in the Universities, among them was "*that the First Cause cannot make many worlds:*".

In 1543, the heliocentric system was introduced by Nicolaus Copernicus (1473 - 1543) in the book *De Revolutionibus Orbium Coelestium.* This implied that our Earth was considered subsequently as a planet among others. In the late sixteenth century, Giordano Bruno (1548 - 1600) defended the plurality of worlds populated by a multiplicity of lifes, as the result of a theological approach.

"*That is how the excellence of God is magnified and the greatness of His empire is demonstrated. He is not glorified in only one Sun, but in countless suns, not in only one Earth and one world, but in thousands of thousands, no, an infinity [of worlds].*"
Giordano Bruno was sentenced

to death and was burned by order of the Church, but nonetheless he was not a martyr for science: his religious opinions were a major factor for the sentence.

In 1609, the discovery of mountains on the Moon, by Galileo Galilei (1564 - 1642), implied that the Moon and the Earth are similar in nature. The Moon is not a perfect sphere. It appeared logical that some men inhabited the Moon as men inhabit Earth. Is some of the relief on the Moon due to men. Satellites around Jupiter suggest that Jupiter is similar to the Earth, so it could have inhabitants. Galileo was much more prudent than Kepler (1571 - 1630) who was very imaginative about his hypotheses. However, Galileo's ideas were only hypotheses…

## 4. Modern epoch

In 1686, Bernard le Bouyer de Fontenelle (1657 - 1757) published in France *Entretiens sur la Pluralité des Mondes* i. e. *A Conversation on the Plurality of Worlds*. This book was a real "best-seller" in the modern sense. It was re-edited many times and translated into many languages. Its influence was very important all through Europe.

"*Stars are so many suns every one of which lights a world*"

Fontenelle assumed that there were inhabitants on the Moon and on every planet, but these inhabitants were different from humans. He spoke also about very various life forms on Earth, even under extreme conditions. This book was written 76 years after Galileo's observations, and shows how great the progress of astronomy had been since then.

In 1754, the book *Amilec ou la graine d'hommes qui sert à peupler les planètes* (*Amilec or the seed of man which serves to populate the planets*) was published in France, by Charles-François Tiphaigne de la Roche (1722 - 1774). This is a story which describes what corresponds to the Big-Bang and to panspermia. "*…This innumerable multitude of Whirls, of Suns, of habitable Earths, which are the components of this vast Universe, all that was contained in the past in a seed the size of which was scarcely equivalent to that of a pea. Its development was done little by little but it is not yet finished*". This book also described a spirit who progressively seeds all the planets. Is it a premonitory description of panspermia? However, it is a legend…

During the 18th century, the existence of life on other planets was commonly accepted by astronomers. We give as an example a sentence in the book of Étienne Bonnot de Condillac (1715 - 1780), which was the course of study for the instruction of the prince of Parma.

"*The skies are full of luminous bodies which, like our Sun, probably make some planets moving on various orbits*"

Condillac, *Cours d'étude pour l'Instruction du prince de Parme* (1775)

Condillac added, in the second part of this sentence:

"*…and the Universe is an immense space where there are no deserted places. Our imagination is as embarrassed to give it limits as not.*"

During the 19th century, the debate took a new turn with the "*discovery*" of Mars canals by Giovanni Virgino Schiaparelli (1835 -1910) during the 1877 and later Mars oppositions (see for example Schiaparelli 1882a et 1882b).

## 5. 20th century

A paper in the French magazine of popular science, *La Nature,* in 1935, *Life in Universe* by Ary J. Sternfeld (1905 - 1980) appears surprisingly modern. Probably the oldest use and definition of Astrobiology are present in this article :

"*The development of natural and astronomical sciences leads to the birth of a science whose object is the habitability of other worlds, astrobiology.*" Sternfeld was a specialist in astronautics. A more detailed study of this pioneer and this article is to be published (Briot, 2012a).

In 1941, six years later, in a article entitled *Astrobiology*, Lafleur defined astrobiology as "*the consideration of life in the universe elsewhere than on Earth*". It is ironic that Sternfeld means Star Field in German and in Yiddish, and Lafleur means The Flower in French ! Between 1945 and 1960, the problem of search for life in the Universe was thoroughly treated at Alma-Ata (Kazakhstan, Soviet Union), in an observatory created and directed by the Russian astronomer Gavriil Adrianovich Tikhov (1875 - 1960). This observatory focused on Astrobotany (this neologism was coined by Tikhov) in order to research possible vegetable life on Mars. A team of botanists studied the spectra of plants growing in conditions as close as possible to conditions on Mars, to determine if these plants, or similar ones, exist on Mars. Unfortunately, the laboratory was suppressed after the death of Tikhov, the new director not believing in life on Mars or on other planets. Because of the separation and lack of communication existing then between the Soviet bloc and other countries, the work of Tikhov remains to this day largely unrecognized (Briot, 2012b).

6. Importance of astrobiology in scientific literature during the 20th century

A good source of information about the importance given to the science of Astrobiology at various times can be found in the books of astronomical bibliography edited at these times. The astronomy bibliographic books Astronomische Jahresbericht were published from 1899 until 1968 by the Astronomisches Rechen-Institut Heidelberg. The language was German. It is a good index of how important the various sectors of astronomy were considered at different times.

In the volumes of Astronomische Jahresbericht, the word "astrobiology" appeared as a keyword as early as 1953, but only in this year, whereas the word "astrobotany" appeared as a keyword in 1949, and again from 1951 to 1954. Papers about astrobotany were mostly written by Tikhov or by scientists from the astrobotany section in Alma-Ata. The phrase "Leben in Kosmos", i.e. "Life in the Universe", was a keyword as early as 1949, but these papers were classified under the heading "Schriften allgemeiner Art" which corresponds to Miscellanea, like "Astronomy and Philately" for example. At the end of the forties, and during the fifties, the papers concerning this subject were still rare but slowly became more and more numerous. Many papers came from the Soviet Union or related countries. Let us note that life and vegetation on Mars was the subject of many papers and their number decreased when it appeared more and more clearly that there is no vegetation on Mars. In 1959, the item "Leben im Universe" gained some importance and became a rubric, instead of being classified indiscriminately under "Schriften allgemeiner Art". In 1969, the volume Astronomische Jahresbericht were replaced by Astronomy and Astrophysics Abstracts, published in English, which lasted until the year 2000. Surprisingly, papers corresponding to the search of extraterrestrial life were debased: they were classified as "Miscellanea". The keywords "Extraterrestrial intelligence" and "Extraterrestrial life" could be found in the Author and Subject Index volumes. None of the words "astrobiology", "bioastronomy" or "exobiology" appeared among the keywords, even in the last volume of Astronomy and Astrophysics Abstracts in 2000. Astrobiology was really treated as a marginal science. However papers about extraterrestrial life became more and more numerous and were often classified under scientific headings. Surpringly, the science Astrobiology remains nearly on the fringe of the mainstream of science up to the year 2000.

7. Conclusion

Because the question: "Is there life elsewhere in the Universe ? " is a fundamental question for humanity, the science which is now named Astrobiology is not a science as young as is generally thought. The history of this science is very old, and is dependent on the scientific results of those times. So this history is not linear and because of that deserves to be carefully studied. We have to emphasize that the real and much more exciting history of astrobiology will begin really when and if we-get a positive answer to the above question. Only then-will the studies of extraterrestrial life be of different forms, be oriented to a precise purpose, and lead to really new results.

Acknowledgements
Thanks are due to my brother Dr. Alain Briot who introduced me to the works of Tiphaine de la Roche and Tikhov. I am also very much indebted to Florence Raulin for many helpful discussions and her precious encouragement, as well as to Roger Hewins and Brigitte Zanda for greatly improving the manuscript.

References

Briot, D. (2012a) Is it the first use of the word Astrobiology ? *to be published in "Astrobiology"*

Briot, D. (2012b) The creator of astrobotany, G.A. Tikhov. In *Astrobiology, History, and Society: Life Beyond Earth and the Impact of Discovery*, ed. Douglas A. Vakoch. Heidelberg: Springer.

Condillac, E. Bonnot de (1775) *Cours d'étude pour l'instruction du prince de Parme*, T. 8, *L'art de raisonner*. Parma: Imprimerie Royale.

Copernicus, N. (1543) *De Revolutionibus Orbium Coelestium*. Nuremberg: Johan Petreius,

Crowe M. J. (2008) *The extraterrestrial life debate, antiquity to 1915*. University of Notre Dame Press, Notre Dame, Indiana

Dick, S. J. (1982) *Plurality of Worlds -The origins of the Extraterrestrial Life Debate from Democritus to Kant*. Cambridge: University Press

Fontenelle, B. le Bouyier de. (1686) *Entretien sur la Pluralité des Mondes*. Paris: Vve. C. Blageart

Galilei, G. (1610) *Sidereus Nuuncius*. Venice: Thomas Baglioni,

Lafleur, L. J. (1941) Astrobiology. *Astronomical Society of the Pacific*, Leaflet n° 143, p.133

Lederberg, J. (1960) Exobiology: approaches to life beyond the Earth. *Science* 132:393-400

Mamikunian, G. (1965) in Current aspects of Exobiology. ed. G. Mamikunian & M. H. Briggs, *Pergamon Press*, p.viii

Schiaparelli, G.V. (1882a) On some observations of Saturn and Mars. *The Observatory*, 5: 221-224

Schiaparelli, G. V. (1882b) Découvertes nouvelles sur la planète Mars. *L'Astronomie*, 1: 216-221

Sternfeld, A. J. (1935) La vie dans l'Univers. *La Nature*, Masson et Cie Eds. N°2956, July 1, pp. 1-12

Tiphaine de la Roche, C.-F. (1753) "Amilec ou la graine d'hommes". Ch. Hugéné, Lunéville